\documentclass[usenatbib]{mn2e}

\usepackage{graphicx}

\title[The quadruple system AO\,Vel]{AO\,Vel: The role of multiplicity in the 
development of chemical peculiarities in late B-type stars\thanks{Based on observations obtained
at the European Southern Observatory, Paranal, Chile, ESO programmes 076.D-0169(A) and 072.D-0235(B).}}

\author[J. F. Gonz\'alez et al.]{J. F. Gonz\'alez$^{1}$\thanks{E-mail: fgonzalez@icate-conicet.gob.ar}, 
S. Hubrig$^{2}$, and F. Castelli$^{3}$\\
$^{1}$Instituto de Ciencias Astron\'omicas, de la Tierra y del Espacio, Casilla 467, 5400 San Juan, Argentina\\
$^{2}$Astrophysical Institute Postdam,  An der Sternwarte 16, 14482 Potsdam, Germany\\
$^{3}$Instituto Nazionale di Astrofisica - Osservatorio Astronomico di Trieste, Via Tiepolo 11, I-34131 Trieste, Italy
}

\begin{document}



\maketitle

\label{firstpage}

\begin{abstract}
We present  high-resolution, high signal-to-noise UVES spectra
of AO\,Vel, a quadruple system containing an eclipsing BpSi star.
From these observations we reconstruct the spectra of the individual components and perform
an abundance analysis of all four stellar members.
We found that all components are chemically peculiar with
different abundances patters. In particular, the two less massive stars show typical characteristics
of HgMn stars. The two most massive stars in the system show variable line profiles indicating the presence of chemical spots.
Given the youth of the system and the notable chemical peculiarities of their components,
this system could give important insights in the origin of chemical anomalies.

\end{abstract}


\section{Introduction}

AO\,Vel (=HD\,68826) was classified as BpSi by \citet{bm73} and
was reported as an eclipsing binary by \citet*{cgv95}.
In fact it is one of the only two double-lined eclipsing binaries with
a BpSi component known to date \citep{he04}.
From light-time effect on the times of minima, \citet{cgv95} 
deduced the presence of a third body.

In our previous paper \citep{paI}, using FEROS spectroscopic observations,
we discovered that this system is actually a spectroscopic quadruple system with components 
close to the ZAMS.
The four stars form two close spectroscopic pairs (periods of 1.58 and 4.15 days)
bound gravitationally to each other in a wide eccentric orbit with a period of 41 yr.
In that paper we combined our radial velocity ({\bf RV}) measurements  with the available photometric data
to derive orbital parameters for both binary systems and to calculate
the absolute parameters of the eclipsing system. For the first time, direct determination
of the radius and the mass was obtained for a BpSi star.

In this work we present high-resolution, high signal-to-noise UVES spectra, which
were used to perform an abundance analysis of all four components of this multiple system.
In Sec.\,2 we present the observations and describe the reconstruction of the component spectra.
In Sec.\,3 we analyze the spectral characteristics of each component and present the results of the 
abundance analysis. In the last Section we discuss the main results and the occurrence of
chemical peculiarities in binary and multiple systems.

\section{ Observations and spectra reconstruction}

Three spectra were obtained in service mode with UVES at VLT-UT2 telescope in October 2005.
The spectra have been taken with the 0.4\,arcsec slit for the blue arm and
the 0.3\,arcsec slit for the red arm on three consecutive nights with two different
dichroics to achieve the highest UVES resolution of 110,000 in the red spectral
region and 80,000 in the blue spectral region.
We used exposure times of 20--30 min, obtaining a S/N ratio above 200 in the spectral range 
4000--8000\AA.

\begin{table*}
  \begin{center}
\label{tab.rv}
    \caption[]{Modified Julian dates of UVES observations, orbital phases, measured RVs, and signal-to-noise ratio at
three different wavelength.}
    \begin{tabular}{cccccccccc}
      \hline
      \multicolumn{1}{c}{MJD} & \multicolumn{1}{c}{Phase AB} &
      \multicolumn{1}{c}{Phase CD} &    \multicolumn{1}{c}{RV$_A$}
	&  \multicolumn{1}{c}{RV$_B$}&  \multicolumn{1}{c}{RV$_C$}&  \multicolumn{1}{c}{RV$_D$}&
S/N$_{\lambda4200}$ & S/N$_{\lambda5600}$ & S/N$_{\lambda8500}$ \\
      \multicolumn{1}{c}{}  & \multicolumn{1}{c}{} & \multicolumn{1}{c}{}
        & \multicolumn{1}{c}{km\,s$^{-1}$}& \multicolumn{1}{c}{km\,s$^{-1}$}& \multicolumn{1}{c}{km\,s$^{-1}$}
	& \multicolumn{1}{c}{km\,s$^{-1}$} \\ \hline 
53669.3357  & 0.7133  & 0.8722 &  \phantom{-}172 $\pm$ 4 &            -180 $\pm$ 4 & \phantom{-}88.2 $\pm$ 2.2 &           -49.7 $\pm$ 0.9 &250 & 300 & 175  \\
53670.3346  & 0.3436  & 0.1129 &            -149 $\pm$ 3 &  \phantom{-}172 $\pm$ 5 &           -38.6 $\pm$ 2.2 & \phantom{0}88.2 $\pm$ 0.9  &310 & 375 & 190 \\
53671.3523  & 0.9859  & 0.3582 &  \phantom{-0}52 $\pm$ 8 &  \phantom{00}-9 $\pm$ 8 &           -50.2 $\pm$ 4.7 &           102.2 $\pm$ 1.6  &235 & 320 & 180\\
\hline
\end{tabular}
\end{center}
\end{table*}

These spectra are analyzed here along with five FEROS spectra described in the previous paper by \citet{paI}.
To calculate separate spectra for the four components of
the system and to measure their RVs, 
the iterative method described by \citet{gl06} was adapted for the multiple system AO~Vel.
This algorithm  computes the spectra of the individual components
and the RVs iteratively.
In each step the computed spectra are used to remove
the spectral features of all but one component
from the observed spectra. The resulting single-lined
spectra are used to measure the RV of that component and to compute
its spectrum by combining them appropriately.

In these calculations 
we combined the three new UVES spectra with our four
FEROS observations taken out of eclipse. 
Since the resolution and S/N-ratio of the UVES spectra were much higher than for FEROS spectra,
we assigned higher weight to these spectra in the last iterations.

Table\,1 lists  the measured  RVs along with the phases (zero at conjunction) corresponding to the
orbits AB and CD. For the calculation of orbital phases we have applied a time correction
of +0.0236\,d and -0.0337\,d for the system AB and CD respectively, in order to take
into account the light time effect.
The last spectrum in this table  was taken at a phase of partial eclipse of the primary star A.
We have taken into account this fact during the calculations of the component spectra, 
considering that the light contribution 
of A is smaller in that spectrum and therefore the line intensities are expected to differ from
those in the remaining spectra. 
If,  for out-of-eclipse phases, star A  contributes a fraction $l_a$ to the total light of the system, 
and during a phase of partial eclipse the light of this star is reduced by a factor  $f$, then
the apparent intensity of the spectral lines of the star A is reduced by a factor $f/(f\cdot l_a +1-l_a)$,
and that of the remaining components is increased by $1/(f\cdot l_a +1-l_a)$.
These scaling factors were estimated to be, in the case of our third UVES spectrum, 
0.85 and 1.10, respectively.

We note that the amplitude of the RV curves of the system AB might be affected by the asymmetry of
the spectral line profiles, which present variations for several elements (see next Section).
This error source has not been considered in the formal error quoted in Table\,1, since
it is difficult to be estimated without determining the surface distribution of the various 
atomic species giving rise for spectral lines used for RV measurements.

The obtained new RV measurements were in general agreement with the orbits published
in  our previous paper \citep{paI}, 
especially for the eclipsing binary AB, for which some orbital parameters
are fixed by the photometric data. In the case of the pair CD,  the new data 
allowed a significant improvement of its orbital parameters.
The orbital parameters recalculated combining all available RV measurements are listed in Table\,2. 
In Fig.\,1 we present the RV curves for
the  eclipsing binary system AB  and for the less massive system  CD.
In the case of the binary AB, which presents apsidal motion, 
the solid line corresponds to the orbit calculated for the epoch of FEROS
observations and the dotted line to the epoch of UVES observations.

\begin{table}
  \begin{center}
\label{tab.orb}
    \caption[]{Orbital parameters for the spectroscopic binaries AB and CD.}
    \begin{tabular}{lcrcl}
      \hline
      \multicolumn{1}{l}{Parameter} & \multicolumn{1}{c}{Unit} & \multicolumn{3}{c}{Value}  \\ \hline 
$P_{CD}$		&days		&4.14933&$\pm$&0.00004\\
$T_{CD}(conj)$		&MJD		&53300.539&$\pm$&0.007\\
$V_{\gamma CD}$		&km\,s$^{-1}$	&21.5	&$\pm$& 0.4 \\
$K_C$			&km\,s$^{-1}$	&94.6	&$\pm$& 1.2\\
$K_D$			&km\,s$^{-1}$	&103.9	&$\pm$& 0.7\\
$e_{CD}$      		&$R_\odot$      &0.017  &$\pm$& 0.006 \\
$\omega_{CD}$		&deg		&261	&$\pm$& 22 \\
$a_{CD}\sin i_{CD}$	&$R_\odot$	&16.48	&$\pm$& 0.12 \\
$q_{CD}$                &               &0.917  &$\pm$& 0.016\\ 
$M_C\sin^3 i_{CD}$	&$M_\odot$	&1.82	&$\pm$& 0.03\\
$M_D\sin^3 i_{CD}$	&$M_\odot$	&1.67	&$\pm$& 0.05\\
\hline
$V_{\gamma AB}$              &km\,s$^{-1}$   &8.1   &$\pm$& 1.8 \\
$K_A$                   &km\,s$^{-1}$   &167.8   &$\pm$& 3.4\\
$K_B$                   &km\,s$^{-1}$   &184.1  &$\pm$& 3.4\\
$a_{AB}$     &$R_\odot$      &10.99$^a$  &$\pm$& 0.15 \\
$q_{AB}$                &               &0.911  &$\pm$& 0.026\\
$M_A$        &$M_\odot$      &3.68$^a$   &$\pm$& 0.14\\
$M_B$        &$M_\odot$      &3.35$^a$   &$\pm$& 0.14\\
\hline
\end{tabular}
\end{center}
$^a$ Calculated using the photometric orbital inclination $i_{AB}=88.5$ deg.
\end{table}

\begin{figure}\label{fig.rv}
  \begin{center}
    \includegraphics[width=6cm,angle=270 ]{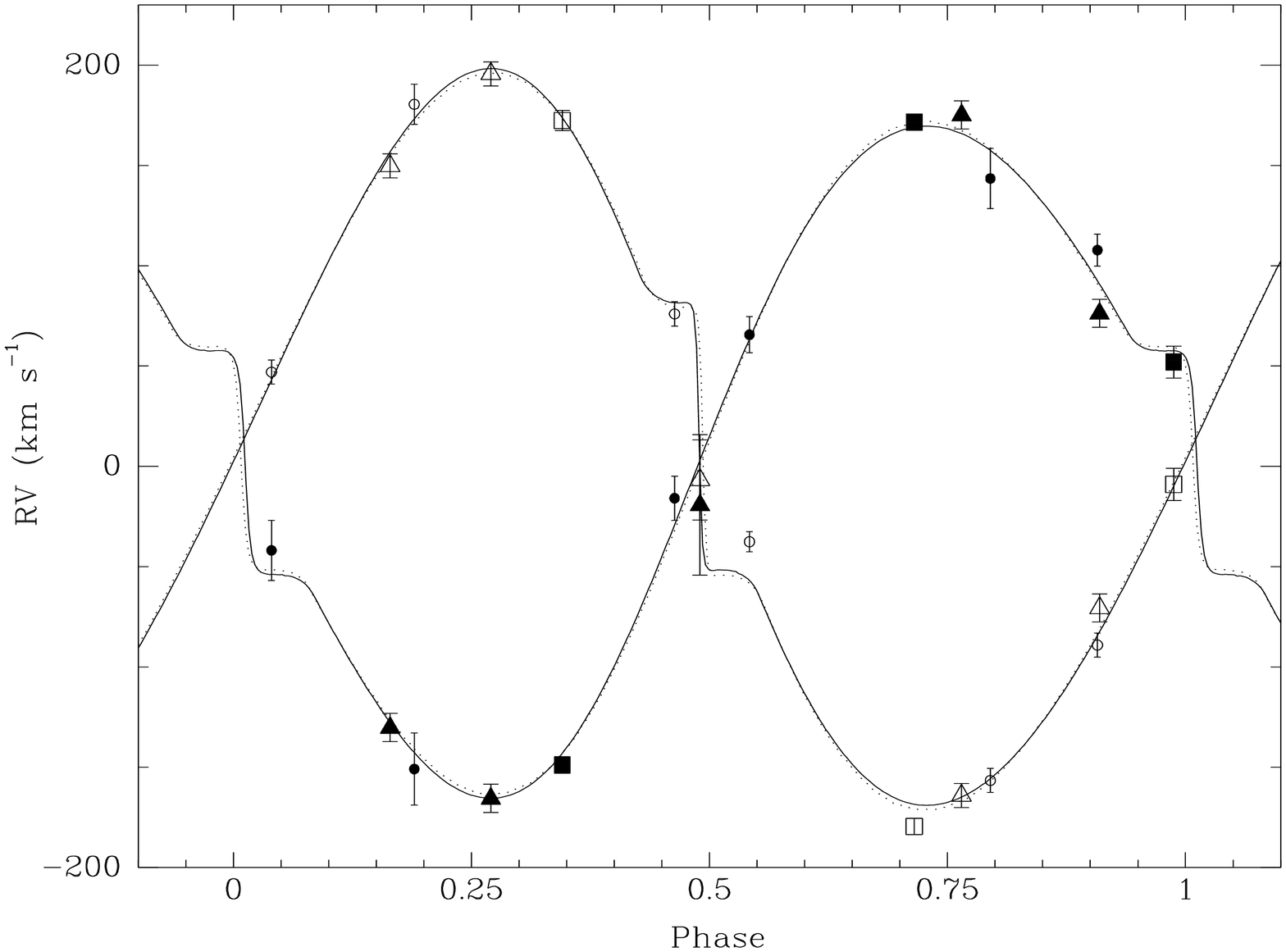}
   \includegraphics[width=6cm,angle=270 ]{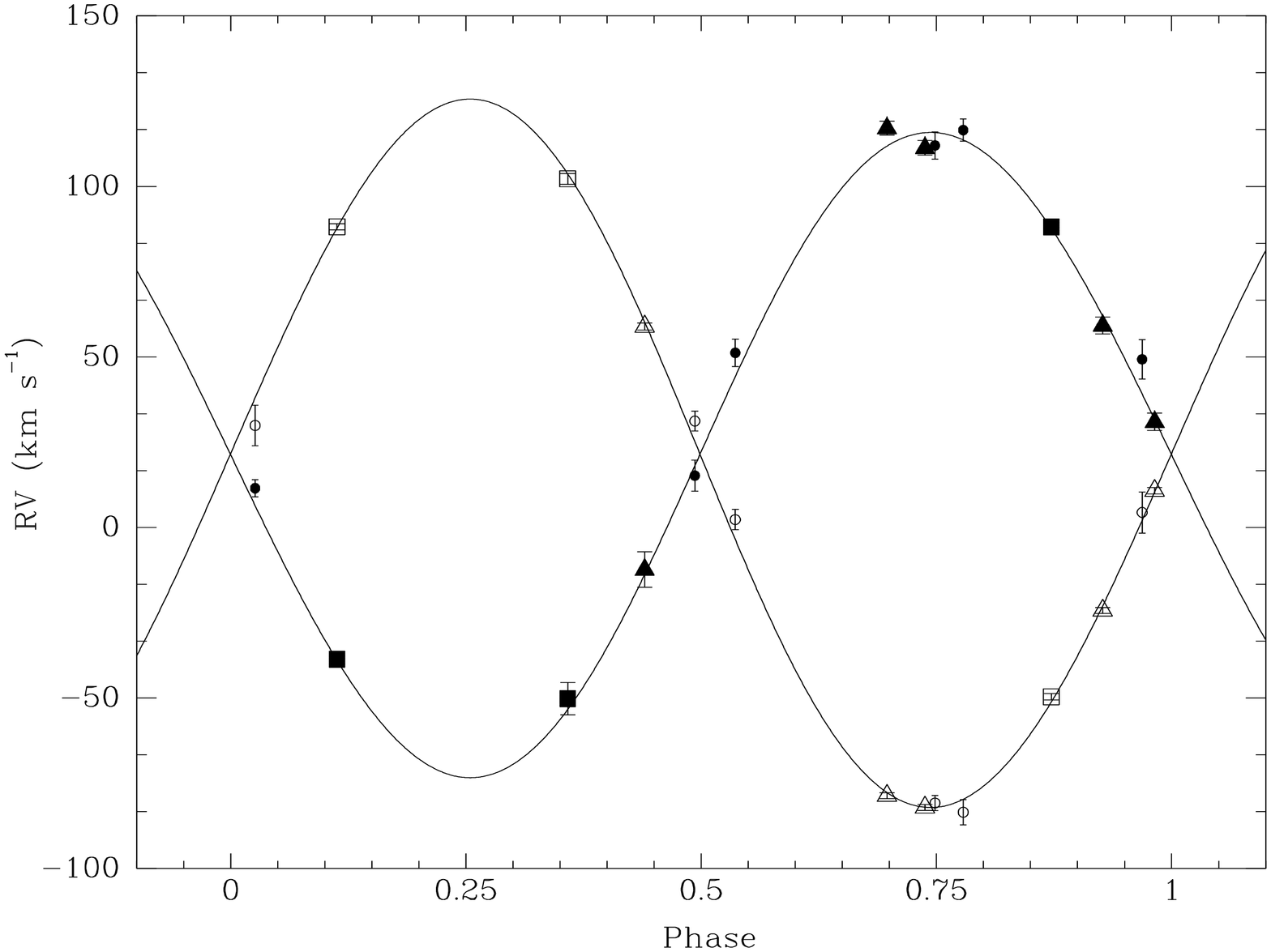}
    \caption{Radial velocity curves for the system AB (upper panel) CD (lower panel). 
Squares, triangles, and small circles are the UVES, FEROS, CASLEO observations, respectively. 
Filled symbols correspond
to the more massive component in each subsystem (A and C).
}
  \end{center}
\end{figure}

Once the spectra of the stellar components have been reconstructed, they must be
scaled to recover the intrinsic intensities of the spectral lines in those stars.
To this aim, as  in our previous paper, we used the photometric relative fluxes given by
\citet{cgv95}.  The relative contribution to the observed continuum near $\lambda5000$
was estimated to be 0.40, 0.28, 0.16, 0.16 for components A, B, C, and D.  
Generally, when applying a spectral disentangling technique, the resulting 
spectra of the individual components 
have a very high quality, since the S/N-ratio increases roughly as the square root of
the number of observed spectra involved in the calculations (usually several tens), 
and diminishes proportionally to the relative flux of the component under consideration (usually close to 0.5).
In the present work a small number of spectra are available for a system of 4 spectral components.
For that reason, the reconstructed spectra, after their continuum has been normalized to one, 
have a S/N-ratio lower than the composite observed spectra. In particular, for the less luminous stars C and D
 the effective S/N resulted between 50 and 80.

\section{Spectral characteristics  and line profile variability}

The primary component of this subsystem is a typical variable BpSi star with
weak He lines.
The spectrum of the component B appears rather normal with strong He and Si lines.
However, a careful inspection of the high-S/N UVES spectra revealed the presence of some  asymmetrical and
variable line profiles in the spectrum of both stars A and B, which 
could be related to the presence of He, Si, and Mg spots.
Since a basic assumption for the reconstruction of the component spectra is the constancy with phase of the
spectral morphology of each star, the disentangled spectra may have small artificial features around 
spectral lines presenting variability. For that reason we preferred, for the analysis of profile variability, to
use only those spectral lines and orbital phases for which the line under consideration is clearly free of
contamination from any other line belonging to the remaining components.
The first two UVES spectra are particularly useful for this purpose since in them both 
binary systems are near opposite quadratures, allowing us to identify clearly the four
spectral components.

In Fig.\,2 we show the  profile of several Si\,\textsc{ii}, Mg\,\textsc{ii},
and He\,\textsc{i} lines, the atomic species that present the most conspicuous lines 
in the spectra of stars A and B.
Both stars A and B exhibit line profile variability.
Figure\,3 shows the spectral region around the Mg\,\textsc{ii} doublet at $\lambda$4481 and
the He\,\textsc{i} line $\lambda$4471.
The upper spectra are the two UVES spectra taken at quadratures.
In the lower spectra, synthetic spectra of C and D have been subtracted from the former.
The Mg\,\textsc{ii} line at $\lambda$4481  shows the same behavior as silicon lines.
In this figure we have plotted the observed spectra to demonstrate that 
 the four components are well resolved, especially star B. 
Consequently, the removal of components
C and D does not affect significantly the shape of the line profiles. As a matter of fact, the profile difference
between the two phases is evident even in the original composite spectra.
In the case of star A,  the lines are blended with those of component
C in one of the two UVES spectra, making somewhat less confident
the  analysis of profile variations.

We note that, even though the reflexion effect due to the mutual irradiation of the components would
distort the line profiles qualitatively in the same way as it is observed for Mg\,\textsc{ii} or Si\,\textsc{ii} lines, 
its expected contribution to the line profile variability is much smaller than the observed variations.
In fact, from the temperature-ratio and the relative separation of the components, we estimate that
the profile variations caused by reflexion effect would be about 5 times smaller than the variations observed in Mg\,\textsc{ii} $\lambda$4481.
Therefore, the cause of these profile variations is not related to proximity effects but 
to asymmetries in the surface chemical distribution.

\begin{figure}\label{fig.prof}
  \begin{center}
    \includegraphics[width=8cm, angle=-90, bb =2 80 557 737]{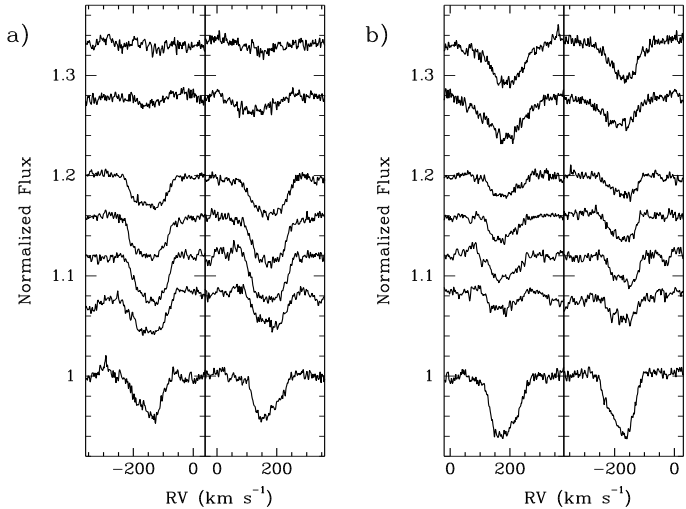}
    \caption{Line profiles variations for stars A (panel a)  and B (panel b) 
near two opposite quadratures: phase 0.34 (left-hand part of each panel) and 0.71 (right-hand
part of each panel).
The spectral line profiles shown are, from top to bottom: He\,\textsc{i} $\lambda\lambda$4471, 4026, Si\,\textsc{ii} 
$\lambda\lambda$6371, 6347, 5056, 5041, and Mg\,\textsc{ii} $\lambda$4481.}
  \end{center}
\end{figure}

\begin{figure}
  \begin{center}\label{fig.prof1}
    \includegraphics[width=8cm]{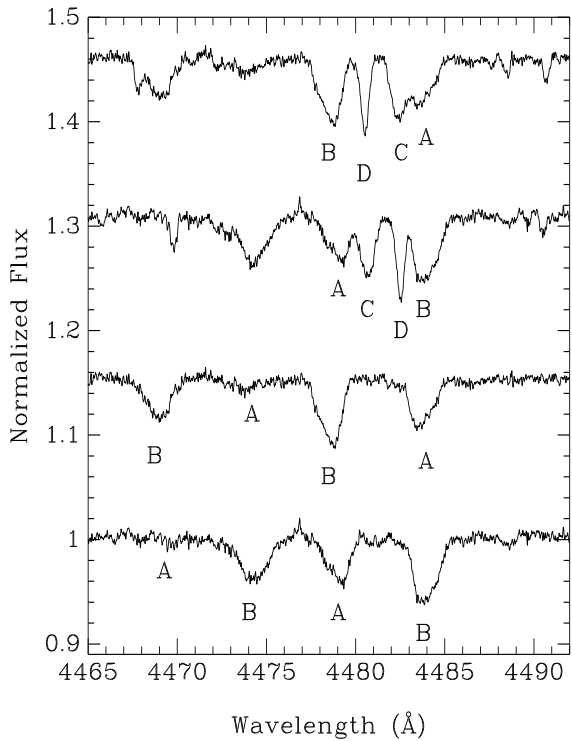}
    \caption{Line profiles variations in components A and B. The two upper spectra are the 
observed UVES spectra for MJD\,53669.3 and 53670.3, respectively. Labels A, B, C, and D mark the position of
the Mg\,\textsc{ii} line $\lambda$4481 for the four components. In the two lower 
spectra C and D components  have been removed. The lines Mg\,\textsc{ii} $\lambda$4481 and
He\,\textsc{i} $\lambda$4471 are marked for components A and B.}
  \end{center}
\end{figure}

Helium lines are weak, especially in star A, so it is not possible to detect clear variations in
the line profiles from the available spectroscopic data.
However, we detect that the strength of He lines at $\lambda$4026 and $\lambda$4471 is variable.
Other strong He\,\textsc{i} lines ($\lambda\lambda$4121, 4922, 5016, 5876) appear blended with Fe or Si
lines belonging to the companions, making  them less useful for the analysis of the line profile variability.

For star B the spectral line profiles of Mg\,\textsc{ii} and Si\,\textsc{ii}
 appear strongly variable, as presented in Figs.\,2 and 3.
Usually, magnesium does not present large equivalent width variations in magnetic peculiar stars \citep*{lcm97}.
However, line profile variability of the magnesium doublet at $\lambda$4481
 has been reported by \citet{k99}, 
who published the first Mg map for an Ap star (CU\,Vir).
They found that, even without showing important  equivalent width  variations,
Mg\,\textsc{ii} $\lambda$4481 
 exhibits notorious profile variations, similar to those observed in AO\,Vel for the components A and B.
The profile shape suggests that Mg is more abundant in the region facing the companion star.
A more complete study of line profiles, that would allow to determine the surface distribution of chemical elements,
will require a larger number of spectra obtained at different phases.

We can assume that the stellar rotation in the short-period  eccentric binary AB is synchronized 
with the angular velocity at periastron, so that the rotational period differs from the orbital period
by a factor 1.163 ($P_{orb}=1.5846$ d, $P_{rot}=1.3622$ d). 
This fact implies that the stellar surface visible at a given orbital phase varies with time,
completing one cycle in 9.7 days, which corresponds to about 6 orbital cycles.
A multi-epoch study of AO\,Vel  would be of great interest to prove whether the location of stellar spots is fixed
on the star surface or is related to the position of the stellar companion.


In our spectra of the C and D components the most interesting, 
and rather unexpected fact is the presence of spectral lines typical of HgMn stars.
The Hg\,\textsc{ii} line $\lambda$3984 is present in both stars C and D, while 
strong Y\,\textsc{ii} and Pt\,\textsc{ii} lines are present 
in the spectrum of component D. 
Although comparatively weak, some Mn\,\textsc{ii} lines have been also identified
in both components.
In the reconstructed spectrum of star D  we identified Pt\,\textsc{ii} lines at 4046 and 4514 
and more than two dozens Y\,\textsc{ii} lines.
No noticeable lines of Zr\,\textsc{ii}, Sc\,\textsc{ii}, Ba\,\textsc{ii}, P\,\textsc{ii}, Xe\,\textsc{ii}, 
Ga\,\textsc{ii} were detected in stars C or D.
The measured central wavelength of Hg\,\textsc{ii} line at 3984, 3983.975\,\AA{} and 3984.08\,\AA{} for
stars C and D, respectively, indicates that heavy Hg isotopes are predominant in these components. 
We will return to this point in the next section.

The weakness of spectral lines for the two less massive stars makes impossible to study line profile
variations in their spectra.

\section{Chemical abundances}

The abundances were determined with the
synthetic spectrum method using ATLAS9 model atmospheres 
and the SYNTHE code \citep{k93}.
Stellar temperatures of companions C and D were estimated from the correlation of 
excitation potentials of Fe lines with abundances \citep*{gnh08}.
Microturbulent velocities were derived from the correlation of equivalent widths of Fe lines with abundances.

In case of stars A and B, however, it was not possible to apply the same procedure since Fe
lines are weaker and they appear broader because of the higher rotation.
Estimated temperatures for these stars were derived from the observed masses and radii, interpolating in
the grid of Geneva theoretical stellar models \citep{geneva} for solar composition.   
Fig.\,4 shows the position of stars A and B in the mass-ratio diagram.
Both stars are located close the ZAMS. 
As is shown in this figure, the masses and radii determined from light and RV curves
correspond to temperatures of 13900$\pm$500\,K and 13200$\pm$450\,K for components A and B, respectively.
Surface gravity were calculated directly from masses and radii.
The adopted atmospheric parameters are listed in Table\,3.

\begin{figure}
  \begin{center}\label{fig.mr}
    \includegraphics[angle=270,width=8cm]{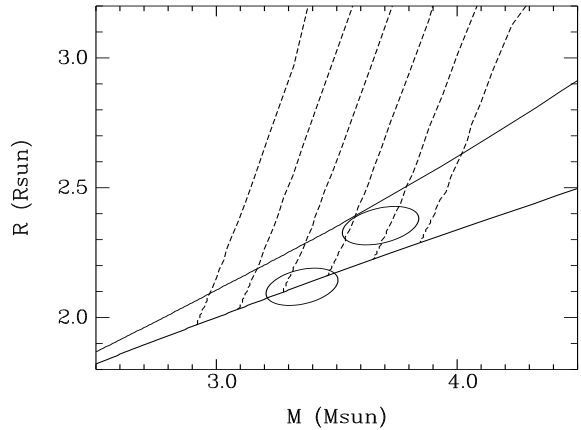}
    \caption{Mass-radius diagram for the components of the more massive binary.           
The parameters of stars A and B, and their error (1-sigma) are represented by the position
and size of the elipses. Continuous lines are the ZAMS and the isochrone log(age)=7.6.
Dashed lines,  from left to right, are isotherms in the range T=12000--14500\,K, with a step of 500\,K.  }
  \end{center}
\end{figure}

The results of the abundance analysis are presented in Table\,4. 
For comparison we list in the last column the solar abundances 
adopted from the work by \citet{sun}.
The abundances of elements not listed in this table were assumed to be solar.
 The complete list of the spectral lines used for abundaces calculation is
included in the Appendix\,A.

\begin{table}
  \begin{center}
\label{tab.atm}
    \caption[]{ Atmospheric parameters}
    \begin{tabular}{ccccc}
      \hline
      \multicolumn{1}{c}{Star} & \multicolumn{1}{c}{T$_{eff}$} &
      \multicolumn{1}{c}{log $g$} &  \multicolumn{1}{c}{$\zeta$} &  \multicolumn{1}{c}{$v \sin i$} \\
      \multicolumn{1}{c}{} & \multicolumn{1}{c}{K} & \multicolumn{1}{c}{} 
	& \multicolumn{1}{c}{km\,s$^{-1}$} & \multicolumn{1}{c}{km\,s$^{-1}$}\\
      \hline \\
A & 13900 $\pm$ 500 & 4.26 $\pm$ 0.03  & 2 & 65 \\
B & 13200 $\pm$ 450 & 4.31 $\pm$ 0.03& 2 & 60 \\
C & 12000 $\pm$ 300 & 4.3 $\pm$ 0.2 & 2 & 40 \\
D & 11500 $\pm$ 300 & 4.2 $\pm$ 0.2 & 1 & 18 \\
\hline
\end{tabular}
\end{center}
\end{table}

\begin{table}
  \begin{center}
\label{tab.abun}
    \caption[]{ Abundances ($\log[N/N(H)]$). Less precise values (error = 0.2--0.4) are marked with a colon. }
    \begin{tabular}{cccccc}
      \hline
      \multicolumn{1}{c}{Element} &  \multicolumn{1}{c}{Star A}&  \multicolumn{1}{c}{Star B}&  \multicolumn{1}{c}{Star C}&  \multicolumn{1}{c}{Star D}&  \multicolumn{1}{c}{Sun} \\
      \hline \\
 He   &   -2.41   &  -1.16    & -2.0:  &$\leq$-3.0:& -1.07  \\ 
 C    &   -4.62   &  -3.87    & -3.5:  & -3.5:    & -3.48 \\ 
 O    &   -3.71   &  -3.56    & -3.7:  & -3.2:    & -3.17 \\ 
 Mg   &   -5.50   &  -4.76    & -4.76  & -4.96    & -4.42 \\
 Si   &   -3.81   &  -4.74    & -4.28  & -4.84    & -4.45 \\
 P    &   -5.7:   &           &        &          & -6.55 \\
 S    &   -4.67   &  -4.67    & -4.67  &          & -4.67 \\
 Ca   &           &           &        &          & -5.69 \\
 Ti   &$\leq$-7.5 &$\leq$-8.0 & -7.0:  & -6.37    & -6.98 \\
 Cr   &           &$\leq$solar& -6.3:  & -5.87    & -6.33 \\
 Mn   &           &           &        & -5.65    & -6.61 \\
 Fe   &   -5.09   &  -4.99    & -4.25  & -4.56    & -4.50 \\
Ni   &           &           &        &$\leq$-6.8& -5.77 \\
 Sr   &           &           &        & -8.27    & -9.03 \\
 Y    &           &           &        & -7.15    & -9.76 \\
 Pt   &           &           &        & -5.94    & -10.2  \\
 Hg   &           &           & -5.05  & -5.70    & -10.9  \\
\hline
\end{tabular}
\end{center}
\end{table}

Star A exhibits an overabundance of Si by 0.64\,dex and underabundance of
He by -1.3$\pm$0.3\,dex.
This explains why He lines are stronger in the star B, even when A is hotter than B
according to the depths of photometric minima.
Star B appears similar to a normal late B-type star, having the 
abundances of helium and silicon close to solar values. However, Fe seems 
to be slightly underabundant in its atmosphere. 
No line of Ti has been observed either in star A or B, indicating that
Ti is underabundant in both stars.
Mg is underabundant in star A by $-1.1$ dex. Similar values are frequently 
observed in magnetic Ap and Bp stars \citep{lcm97}.

We note that the presented abundances in stars A and B should be taken with caution, since  
the detected line profile variations suggest 
a non-uniform chemical distribution of at least a few elements.

The subsystem CD shows abundances which are typical for HgMn stars.
Star C shows a strong overabundance of Hg by 5.8\,dex. 
Other metals like Fe, Ti, Mg, and Si show normal solar abundances in star C.
Abundances of C, O, Ga, Sc, and Sr seem also to be solar. 
Helium lines are only marginally detected in star C and the corresponding abundance
is consequently rather uncertain (about 0.4 dex). It is clear, however, that
He is underabundant in star C. 

Star D exhibits overabundances of Hg by 5.2\,dex,
Y by 2.5\,dex, Pt by 5.2\,dex, and Sr by 0.8\,dex,
relative to the solar values. Ga and Sc lines are not observed. 

The uncertainties of the abundance determination, typically 0.1--0.2, are somewhat larger than in usual
analyses
even though the observed spectra have high S/N-ratio and high resolution.
Three different  reasons contribute to the degradation of the accuracy.
First, the effective S/N-ratio of the normalized spectra of the individual components
is comparatively low, since the line strength is diluted by the continua of the 4 components.
In the case of stars C and D, for example, the S/N is diminished by a factor 6.
Second, the disentangling process using a small number of observed spectra might generate
small fake features, especially close to strong variable lines.
Finally, the uncertainty of the line equivalent widths involves the uncertainty
in the relative light contribution of each component to the total light of the system, i.e.
the normalization scale factors adopted from the photometric data.
We note however that, the  chemical peculiarities mentioned above are 
much larger than the uncertainties, and the classification of stars C and D as
HgMn stars is absolutely out of doubt.

\begin{figure*}
\label{fig.CD}
  \begin{center}
    \includegraphics[bb= 280 58 583 733, angle=270, width=\textwidth]{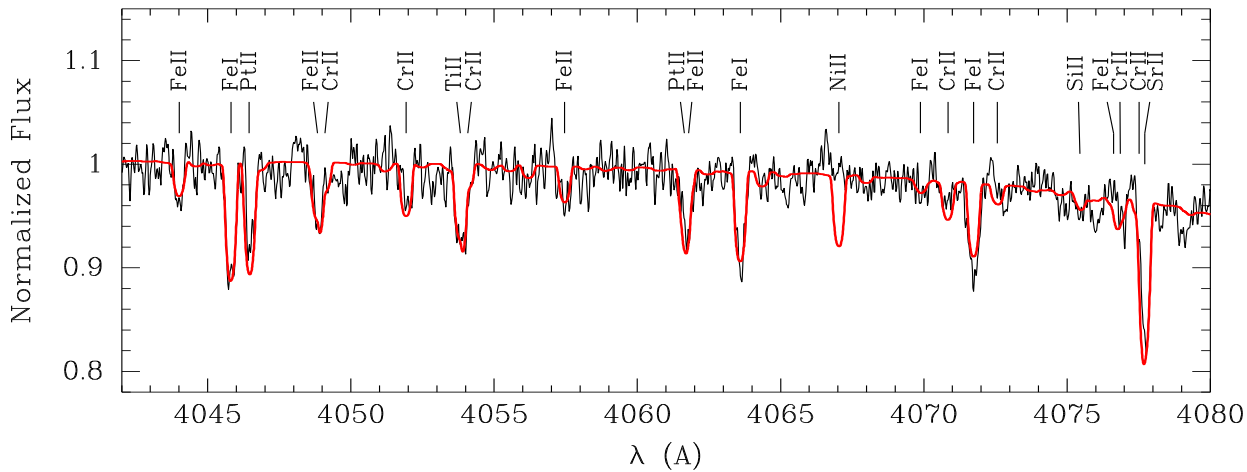}
    \caption{  Comparison of the reconstructed spectrum of the star D  and
 the synthetic spectrum (gray line, red in online version) in the spectral region around Pt and Sr lines. 
  In the calculation of this synthetic spectrum, Ni abundance was assumed to be solar so that
   the large Ni underabundance is better appreciated. }
  \end{center}
\end{figure*}

In Fig.\,5 we present the synthetic spectrum of star D in 
the spectral region containing  Pt\,\textsc{ii}\,4046, Pt\,\textsc{ii}\,4061,
 Sr\,\textsc{ii}\,4077, and several Fe and Cr lines.
As it is visible in this figure, Ni is underabundant in this star.
Fig.\,6 shows the synthetic spectra for the regions of the Hg\,II line at $\lambda3984$.
Even when the rotational broadening is significant, some information about the isotopic
composition can be derived from the reconstructed spectra of components C and D. 
With this aim the line profile was fitted with a synthetic spectrum
in which  Hg is assumed to be a mixture of the lightest (196) and the heaviest (204) isotopes.
As it can be seen in the figure, the best fit suggests that the heaviest isotope is by far
the most abundant in both components of the less massive subsystem.
 
\begin{figure}
  \begin{center}\label{fig.hg}
 \includegraphics[width=0.38\textwidth,angle=270]{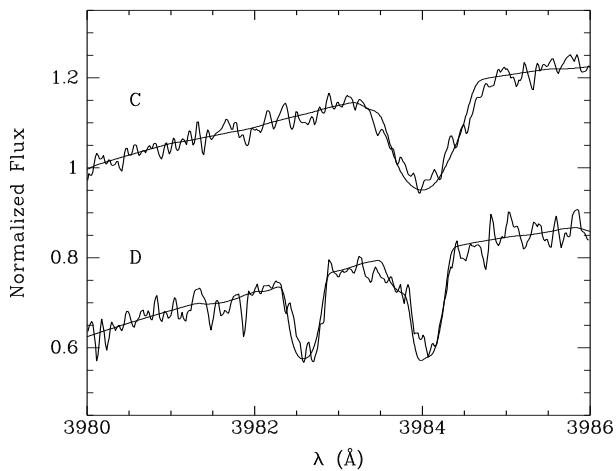}
    \caption{
Observed spectra (thin line) and synthetic spectra (heavy line) for stars C and D
in the region of the Hg\,\textsc{ii} line $\lambda$3984 and the Y\,\textsc{ii} line $\lambda$3982.
The isotopic composition for Hg was assumed to be 196=0.9\%, 204=99.1\% for starC and 196=4\%, 204=96\% for star D. 
}
  \end{center}
\end{figure}

\section{Discussion}

Using high-S/N high-resolution spectra obtained with UVES, 
we have determined chemical abundance for all four components of the multiple
system AO\,Vel. Further, line profile variations were detected in the spectra of
the two most massive stars belonging to the AB system.

Since the multiple system AO\,Vel is formed by four gravitationally bound stars, we can assume that  they
have the same original chemical composition and the same age. All components 
are close to the ZAMS and have already
developed different chemical anomalies in their atmospheres, depending on
their temperature and probably due to the membership in binary systems.
The observed masses and radii for the components of the eclipsing pair
suggest that the system age is less than about 40 million years.

The connection between HgMn peculiarity and membership in binary and multiple systems seems to be
supported by two observational facts. On the one hand there is a high frequency
of binaries and multiples among HgMn stars \citep*{has08a}.
On the other hand, in some binaries the location of chemical spots seems to be related with
the position of the binary companion
\citep[][see also \citealt{shgs09}]{h06a}.
These studies indicates that the presence of a companion can play a significant role
in the development and distribution of chemical anomalies on the surface of HgMn stars,
probably due to proximity effects
on the atmospheric temperature or due to the magnetic field induced in a close binary system.
The role that magnetic fields possibly play in the development of anomalies in
HgMn stars, which mostly appear in binary systems, has
never been critically tested by astrophysical dynamos.
Recent magnetohydrodynamical simulations revealed
a distinct structure for the magnetic field topology similar to the fractured elemental rings
observed on the surface of HgMn stars \citep*[see e.g. Fig.~1 in ][]{hga08b}.  The combination of
differential rotation and a poloidal magnetic field was
studied numerically by the spherical MHD code of \citet{h00}.  The presented typical
patterns of the velocity and the magnetic field on the
surface of the star may as well be an indication for element redistribution on (or in) the star.

Generally He and Si variable Bp stars possess large-scale organized magnetic
fields which in many cases appear to occur essentially under the form
of a single large dipole located close to the centre of the star. The magnetic field is usually diagnosed
through observations of circular polarization in  spectral lines.
\citet{h06b} used the multi-mode instrument FORS\,1 mounted on the 8 m Kueyen telescope 
of the VLT to measure the mean longitudinal magnetic field  in the system AO Vel.  
No detection was achieved with the
single low-resolution measurement  (R=2000) resulting in $\left<B_{\rm z}\right>$\,=\, 80$\pm$64\,G.
We should, however, keep in mind that the observed spectropolarimetric spectrum is a composite spectrum consisting
of four spectropolarimetric spectra belonging to A, B, C and D components. Each of the companions can possess an
individual  magnetic field of different geometry,  strength and polarity.
Since the inhomogeneous elemental distribution of Si and Mg is also observed on the surface of the component B,
the presence of a large-scale organized magnetic field is quite possible in this component too.
The components C and D show the peculiarity character typical of HgMn stars.  \citet{h06b} showed that
longitudinal magnetic fields in  HgMn stars are rather weak, of the order of a few hundred Gauss and less, and
the structure of their fields must be sufficiently tangled so that it does not produce a strong net observable
circular polarization signature. 

Thus, the non-detection of a magnetic field in the system AO Vel can be possibly explained  by  the
dilution of the magnetic  signal due to the superposition  of four differently polarized  spectra.
On the other hand, the magnetic field of each component should be detectable
with high resolution spectropolarimeters making use of the Zeeman effect in individual metal lines.

The time scale for the peculiarities to be developed remain unclear since 
the results of studies  of evolutionary status of chemically peculiar stars of  different type 
show somewhat contradictory results. 
The early work by \citet{a79} suggested that the frequency of peculiar stars of Si 
and HgMn groups increases with age and particularly there exist no peculiar star on the ZAMS.
In a more recent photometric search for peculiar stars in five young clusters, \citet{ppm02} 
also concluded that the CP phenomenon needs at least several Myr to start being effective.
On the other hand, some chemically peculiar stars has been detected within very
young associations or even in  star forming regions like Ori OB1 \citep{al77,wl99},
Lupus 3 \citep{ch07}, and L988 \citep{hd06}.  
The results of the present paper show the existence of coeval stars with chemical peculiarities of
Si and HgMn types, close to the ZAMS indicating that the age threshold for these peculiarities is similar 
for both subgroups of chemically peculiar stars.
Our study supports the idea that these  chemical peculiarities originate quite soon after the 
star formation.

It is noteworthy, that the subsystem CD, the  less massive binary of the AO\,Vel, shows several similarities with the
eclipsing binary AR\,Aur belonging to the Aur OB1 association. 
Both systems are formed by stellar components  very close to the ZAMS or 
even in the pre-MS stage \citep{nj94,paI}, and they
have almost the same orbital period, 4$^d$.13  and 4$^d$.15  for AR\,Aur and AO\,Vel\,CD, respectively.
In both systems the primary is a HgMn star, but while 
in the AO\,Vel CD system the secondary companion shares the same peculiarity, in AR\,Aur the secondary
is a normal star \citep{k95}.
In fact, the primary component of AR\,Aur should be compared with 
star D of the system AO\,Vel, since they have a similar effective temperature.
This similarity in the physical and orbital properties seems to have been translated into
the development of similar pattern of chemical abundances:
notorious overabundance of Hg and Pt, overabundance  of Sr, Y, and Mn at the level of 1--3 dex, slight
overabundance of Cr and Ti, and the abundance of Si and Mg close to the solar abundance, or slightly subsolar.
We selected a few additional HgMn spectroscopic binaries with effective temperatures close to 11,000\,K:
HD\,32964, HD\,89822,  HD\,173524, and HD\,191110.
In Fig.\,7  the abundances of various elements in AR\,Aur and selected binaries are compared
with those of the AO\,Vel D companion. 
The abundances are presented for HgMn stars member of double-lined spectroscopic binaries with 
periods between 4 and 12 days and having effective temperatures in the range 10\,700--11\,700\,K.
For HD\,173524 and HD\,191110 we plotted the abundances determined in both components.
The abundances values have been taken from \citet{a94}, \citet{cl04}, and \citet*{cll03}.
It is remarkable, how much similar the chemical composition of these stars is.
We note, however, that the abundance distribution on the stellar surface of HgMn stars 
is probably inhomogeneous, and consequently, 
the chemical composition derived from spectroscopic observations that do not 
cover the rotational cycle, should be taken just as indicative values until accurate 
abundance values obtained using Doppler imaging technique become available.
The spotted character of magnetic Bp-Ap stars is well known, but also in the
case of HgMn stars the presence of chemical spots or belts on the surface is not
uncommon \citep{hga08b,shgs09}.

\begin{figure}
  \begin{center}\label{fig.abun}
 \includegraphics[ height=8cm, angle=270]{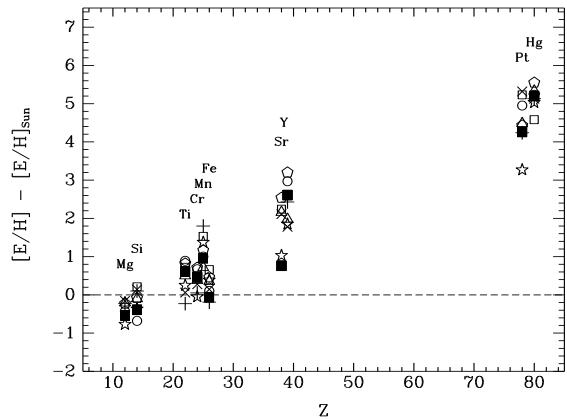}
    \caption{
 Abundances of the AO\,Vel D companion (filled squares) compared with published abundances for another 
seven binary components with similar effective temperatures:
HD\,32964B (circles), HD\,34364A (open squares), HD\,89822A (pentagons), HD\,173524A (stars),  
HD\,173524B (triangles), HD\,191110A (plusses), and HD\,191110B (crosses).
The dashed line represents the solar abundance.}
  \end{center}
\end{figure}

As already mentioned above, the atmospheric chemical composition of AR\,Aur exhibits a non-uniform surface distribution with
a very interesting pattern related to the position of the companion \citep{h06a}.
A similar study in AO\,Vel would be worthwhile to verify whether the elemental distribution on the 
stellar surfaces of C and D components show a similar behavior.
Furthermore, an intensive spectroscopic campaign that would allow the reconstruction
of chemical maps for all four components should provide important information
for the proper understanding of the origin of chemically peculiar stars. \\

\appendix

\section{List of spectral lines}

Table\,\ref{tab.lines} lists all the spectral lines used
for abundance determination.
The lines marked with an asterisk are blends. The abundance was determined using the
fitting by a synthetic spectrum.

\newcommand{\teff}{$T_{\rm eff}$}
\newcommand{\logg}{$\log~g$}

\pagestyle{empty}

\begin{table*}[]\label{tab.lines}
\begin{flushleft}
\caption{Spectral lines used for abundace determination.}
\begin{tabular}{lllllllll}
\noalign{\smallskip}\hline
\multicolumn{1}{c}{ion}  &
\multicolumn{1}{c}{$\lambda$(\AA)}&
\multicolumn{1}{c}{$\log$\,gf}&
\multicolumn{1}{c}{Source}&
\multicolumn{4}{c}{$\log\,N/H$}\\
\noalign{\smallskip}\hline
       &       &  &   &
\multicolumn{1}{c}{A}  &
\multicolumn{1}{c}{B}  &
\multicolumn{1}{c}{C}  &
\multicolumn{1}{c}{D}  &
\\
\noalign{\smallskip}\hline
He\,I       & 4026.1844$^{*}$ &$-$2.625&NIST3& $-$2.3  & $-$1.2  & $-$2.0 & $\le$$-$3.0 \\
            & 4026.1859$^{*}$ &$-$1.448&NIST3&\\
            & 4026.1860$^{*}$ &$-$0.701&NIST3&\\
            & 4026.1968$^{*}$ &$-$1.449&NIST3&\\
            & 4026.1983$^{*}$ &$-$0.972&NIST3&\\
            & 4026.3570$^{*}$ &$-$1.324&NIST3&\\
            & 4387.9291       &$-$0.883&NIST3& $--$    & $-$1.2  & $--$ &$--$\\
            & 4471.4704$^{*}$ &$-$2.203&NIST3& $-$3.0  & $-$1.25  & $-$2.0 & $\le$$-$3.0\\
            & 4471.4741$^{*}$ &$-$1.026&NIST3\\
            & 4471.4743$^{*}$ &$-$0.278&NIST3\\
            & 4471.4856$^{*}$ &$-$1.025&NIST3\\
            & 4471.4893$^{*}$ &$-$0.550&NIST3\\
            & 4471.6832$^{*}$ &$-$0.903&NIST3\\
            & 4921.9310       &$-$0.435&NIST3&  $-$1.8 & $-$1.2 & $--$ &$--$\\
            & 5015.6776       &$-$0.820&NIST3 &$--$ &$-$1.0 & $--$ &$--$\\
            & 5875.5987$^{*}$ &$-$1.516&NIST3 & $-$2.7   & $-$0.95 & $\le$$-$2.4 &$-$2.0\\
            & 5875.6139$^{*}$ &$-$0.341&NIST3\\
            & 5875.6148$^{*}$ &$+$0.408&NIST3\\
            & 5875.6251$^{*}$ &$-$0.340&NIST3\\
            & 5875.6403$^{*}$ &$+$0.137&NIST3\\
            & 5875.9663$^{*}$ &$-$0.215&NIST3\\
            & 6678.1517&$+$0.329&NIST3& $-$2.2   & $-$0.9 & $\le$$-$3.0 &$-$1.8\\
\\
C\,II  & 4267.001$^{*}$&$+$0.563&NIST3 & $-$4.62        & $-$3.87 & $-$3.52 & $-$3.52\\
       & 4267.261$^{*}$&$+$0.716&NIST3\\
       & 4267.261$^{*}$&$-$0.584&NIST3\\
\\
O\,I  & 6155.961$^{*}$&$-$1.363& NIST3 &$-$3.71    & $-$3.56 & $-$3.71 & $-$3.21\\
      & 6155.971$^{*}$&$-$1.011& NIST3 &\\
      & 6155.989$^{*}$&$-$1.120& NIST3\\
      & 6156.737$^{*}$&$-$1.487& NIST3 &$-$3.71    & $-$3.56 & $-$3.71 & $-$3.21\\
      & 6156.755$^{*}$&$-$0.898 &NIST3\\
      & 6156.778$^{*}$&$-$0.694 &NIST3\\
      & 6158.149$^{*}$&$-$1.841& NIST3 &$-$3.71    & $-$3.56 & $-$3.71 & $-$3.21\\
      & 6158.172$^{*}$&$-$0.995& NIST3\\
      & 6158.187$^{*}$&$-$0.409& NIST3\\
\\
Mg\,II & 4481.126$^{*}$& $+$0.749&NIST3 & $-$5.5    & $-$4.76     &$-$4.76  & $-$4.96\\
       & 4481.150$^{*}$& $-$0.553& NIST3\\
       & 4481.325$^{*}$& $+$0.594& NIST3\\
\\
Si\,II & 3853.665 & $-$1.341& NIST3 & $-$4.4    & $-$4.3   & $-$5.6 &$-$5.0\\
       & 3856.018 & $-$0.406& NIST3 & $-$4.4    & $-$4.9   & $-$5.6 &$-$5.0\\
       & 3862.595 & $-$0.757 &NIST3  & $-$4.0    & $-$4.9   & $-$5.6 &$-$5.0\\
       & 3954.300$^{*}$ & $-$1.040 &KP & $-$3.6 &$--$ & $--$ &$--$\\
       & 3954.504$^{*}$ & $-$0.880 &KP\\
       & 4128.054 & $+$0.359 &NIST3 & $-$4.1    & $-$4.8  & $-$4.3 & $-$5.2\\
       & 4130.872$^{*}$ & $-$0.783 &NIST3 & $-$4.1    & $-$4.8  & $-$4.3   & $-$5.0\\
       & 4130.894$^{*}$ & $+$0.552 &NIST3\\
       & 4187.128$^{*}$ & $-$1.050 &KP & $-$3.7 &$--$   & $--$ &$--$\\
       & 4187.128$^{*}$ & $-$2.590 & KP\\
       & 4187.151$^{*}$ & $-$1.160 & KP\\
       & 4190.724 & $-$0.351 & LA & $-$3.7 & $--$ &$--$ &$--$\\
       & 4198.133 & $-$0.611 & LA & $-$3.7 & $--$ &$--$ &$--$\\
       & 4200.658$^{*}$ & $-$0.889 & NIST3 & $-$3.7&$--$&$--$& $--$\\
       & 4200.887$^{*}$ & $-$2.034&NIST3\\
       & 4200.898$^{*}$ & $-$0.733&NIST3\\
       & 5041.024 & $+$0.029 & NIST3& $-$3.8    & $-$4.6 & $-$3.9  & $-$4.5\\
\hline
\noalign{\smallskip}
\end{tabular}
\end{flushleft}
\end{table*}

\begin{table*}[]
\begin{flushleft}
\contcaption{}
\begin{tabular}{lllllllll}
\noalign{\smallskip}\hline
\multicolumn{1}{c}{ion}  &
\multicolumn{1}{c}{$\lambda$(\AA)}&
\multicolumn{1}{c}{$\log$\,gf}&
\multicolumn{1}{c}{Source}&
\multicolumn{4}{c}{$\log\,N/N_H$}\\
\noalign{\smallskip}\hline
       &       &  &   &
\multicolumn{1}{c}{A}  &
\multicolumn{1}{c}{B}  &
\multicolumn{1}{c}{C}  &
\multicolumn{1}{c}{D}  &
\\
\noalign{\smallskip}\hline
       & 5055.984$^{*}$&$+$0.523&NIST3& $-$4.0    & $-$4.7 & $-$4.3 & $-$4.8\\
       & 5056.317$^{*}$&$-$0.492&NIST3\\
       & 5185.232$^{*}$& $+$0.472& LA& $-$3.8    & $--$ & $--$ & $--$\\
       & 5185.520$^{*}$&$-$1.603& NIST3\\
       & 5185.520$^{*}$&$-$0.302& NIST3\\
       & 5185.555$^{*}$&$-$0.456& NIST3\\
       & 5688.817 & $+$0.126 & NIST3& $-$3.6    & $--$ & $--$ &$--$\\
       & 5701.370 & $-$0.057 & NIST3& $-$3.6    & $--$ & $--$ &$--$\\
       & 5957.559 &$-$0.225 & NIST3& $-$4.0    & $-$4.9 & $-$4.6 & $-$4.8\\
       & 5978.930 &$+$0.084& NIST3& $-$4.0    & $-$5.4 & $-$4.6 & $-$4.8\\
       & 6347.103&$+$0.149&NIST3& $-$3.6    & $-$4.4 & $-$4.0 & $-$4.6\\
       & 6371.371 &$-$0.082&NIST3& $-$3.7    & $-$4.5 & $-$4.3 & $-$4.6\\
\\
P\,II  & 4178.463& $-$0.409 &HI &  $-$6.59  & $--$ & $--$ &$--$\\
       & 6024.178&$+$0.137& NIST3   &$-$5.39 & $--$ & $--$&$--$\\
       & 6034.039&$-$0.220& NIST3   &$-$5.39  &$--$ & $--$&$--$\\
       & 6043.084&$+$0.416& NIST3   &$-$5.39  &$--$ & $--$&$--$\\
\\
S\,II &  4153.068 & $+$0.617& NIST3 & $-$4.71 &$-$4.71 & $-$4.71 &$-$4.71\\
      &  5027.203 & $-$0.705& NIST3 & $-$4.71 &$-$4.71 & $-$4.71 &$-$4.71\\

\\
Ti\,II  &4399.765&$-$1.190& PTP&$\le$$-$7.5 &$\le$$-$8.0&$-$7.02    &$--$\\
        &4411.072&$-$0.670& PTP&$\le$$-$7.5 &$\le$$-$8.0&$-$7.02    &$-$6.37\\
        &4418.714&$-$1.970& PTP&$\le$$-$7.5 &$\le$$-$8.0&$-$7.02    &$-$6.37\\
        &4468.492&$-$0.620& NIST3&$\le$$-$7.5 &$\le$$-$8.0 &$-$7.02    & $-$6.37\\
\\
Cr\,II & 4261.913&$-$1.531& K03Cr&$--$  &$--$ &$-$6.37 & $-$5.87 \\
       & 5237.329&$-$1.160& NIST3&$--$  &$--$ &$-$6.37 & $-$5.87\\
       & 5313.590&$-$1.650& NIST3&$--$  &$--$ &$-$6.37 & $-$5.87\\
\\
Mn\,II & 4206.367 &$-$1.590& K03Mn&$--$ &$--$ &$-$5.65 & $-$5.65\\
       & 4478.635 &$-$0.942& K03Mn&$--$ &$--$ &$-$5.65 & $-$5.65\\
\\
Fe\,II& 4178.862 &$-$2.440& FW06&  $-$5.75  & $-$5.0 &$--$  &$--$\\
      & 4233.172&$-$1.809 & FW06&  $--$     & $--$   & $-$4.24 & $-$4.54\\
       & 4303.176&$-$2.610& FW06&  $-$5.75 & $-$5.0 & $--$ &$--$\\
       & 4385.387&$-$2.580& FW06  & $-$5.75 & $-$5.0:& $--$&$--$\\
       & 4416.830&$-$2.600& FW06 & $-$5.75 & $-$5.0 & $-$3.50 & $-$4.14\\
       & 4923.927&$-$1.210& FW06&  $-$5.75 & $-$5.3 & $-$4.24 & $-$4.85\\
       & 5018.440&$-$1.350& FW06 & $-$4.90 & $-$4.8 & $-$4.00 & $-$4.54\\
       & 5100.607$^{*}$&$+$0.144& K09 &$--$  &$--$     & $-$4.24 & $-$4.54\\
       & 5100.734$^{*}$&$+$0.671& J07&\\
       & 5169.033& $-$0.870& FW06 & $-$5.20 & $-$4.8 & $-$4.24 & $-$4.54\\
        & 5197.577&$-$2.054 &FW06 & $--$        &$--$        & $-$4.34 & $-$4.70\\
       & 5227.483&$+$0.831 &J07 &  $--$      & $--$       & $-$4.24 & $-$4.54\\
       & 5234.625&$-$2.210& FW06 & $--$  & $--$       & $-$4.34 & $-$4.54\\
       & 5276.002 &$-$1.900&FW06& $-$4.8 &  $-$5.8  & $-$4.54 & $-$4.74\\
       & 5316.615&$-$1.780& FW06 & $-$4.8 & $-$5.0 & $-$4.34 & $-$4.74\\
       & 5362.741$^{*}$&$-$0.708&K09 &$--$ & $--$        & $-$4.00 &$-$4.34\\
       & 5362.869$^{*}$&$-$2.855& K09 &\\
       & 5362.967$^{*}$&$+$0.008&K09\\
\\
Ni\,II  & 4015.474&$-$2.410&K03Ni &$--$ &$--$ &$--$ & $\le$$-$6.79 \\
       & 4067.031&$-$1.834&K03Ni  &$--$ &$--$ &$--$ & $\le$$-$6.79 \\
\\
Sr\,II & 4077.709&$+$0.151& NIST3&$--$ &$--$ &$--$ & $-$8.27\\

\hline
\noalign{\smallskip}
\end{tabular}
\end{flushleft}
\end{table*}

\begin{table*}[]
\begin{flushleft}
\contcaption{}
\begin{tabular}{lllllllll}
\noalign{\smallskip}\hline
\multicolumn{1}{c}{ion}  &
\multicolumn{1}{c}{$\lambda$(\AA)}&
\multicolumn{1}{c}{$\log$\,gf}&
\multicolumn{1}{c}{Source}&
\multicolumn{4}{c}{$\log\,N/N_H$}\\
\noalign{\smallskip}\hline
       &       &  &   &
\multicolumn{1}{c}{A}  &
\multicolumn{1}{c}{B}  &
\multicolumn{1}{c}{C}  &
\multicolumn{1}{c}{D}  &
\\
\noalign{\smallskip}\hline

Y\,II  & 3982.592&$-$0.493&NIST3&$--$ &$--$ &$--$ & $-$7.45\\
       & 4235.727&$-$1.509&NIST3&$--$ &$--$ &$--$ & $-$6.60\\
       & 4309.620&$-$0.745&NIST3&$--$ &$--$ &$--$ & $-$6.95\\
       & 4883.684&$+$0.071&NIST3&$--$ &$--$ &$--$ & $-$7.25\\
       & 4900.120&$-$0.090&NIST3&$--$ &$--$ &$--$ & $-$7.25\\
       & 5200.406&$-$0.579&NIST3&$--$ &$--$ &$--$ & $-$7.45\\
       & 5205.722&$-$0.342&NIST3&$--$ &$--$ &$--$ & $-$7.15\\
\\
Pt\,II & 4046.443&$-$1.190&ENG &$--$ &$--$ &$--$ & $-$5.94\\
       & 4061.644&$-$1.890&ENG &$--$ &$--$ &$--$ & $-$5.94\\
       & 4288.371&$-$1.570&ENG &$--$ &$--$ &$--$ & $-$5.94:\\
\\
Hg\,II & 3983.890&$-$1.510&NIST3&$--$&$--$ &$--$ & $-$6.55\\
\hline
\noalign{\smallskip}
\end{tabular}

NIST3: NIST atomic spectra Database, version 3 at http://physics.nist.gov;\\
FW06 : \citet{fu06} \\ 
PTP: \citet{ptp02} \\
LA : \citet{la85} \\
KP : \citet{kp75} \\
HI : \citet{Hi88} \\ 
J07: Johannson (2007), private communication\\
K03Cr:Kurucz (2003), http://kurucz.harvard.edu/atoms/2401/gf2401.pos\\
K03Mn:Kurucz (2003), http://kurucz.harvard.edu/atoms/2501/gf2501.pos\\
K03Ni:Kurucz (2003), http://kurucz.harvard.edu/atoms/2801/gf2801.pos\\
K09: Kurucz (2009), http://kurucz.harvard.edu/atoms/2601/gf2601.pos\\
\end{flushleft}
\end{table*}

We thank to Charles R. Cowley and Rainer Arlt for helpful discussions.

{}

\end{document}